\documentclass{emulateapj}
\usepackage{natbib}
\usepackage{epsfig}
\usepackage{graphicx}
\usepackage{subfigure}
\usepackage{float}
\usepackage{amsmath}
\usepackage{color}
\usepackage{mathtools}
\usepackage{amssymb}
\usepackage{amsfonts}
\usepackage{units}
\usepackage[colorlinks,linkcolor=blue,anchorcolor=green,citecolor=blue]{hyperref}
\def\asn{ASASSN-15lh}

\shorttitle{X-ray emission from ASASSN-15lh}
\shortauthors{ Huang \& Li}

\begin{document}

\title{Persistent X-ray emission from ASASSN-15lh: massive ejecta and pre-SLSN dense wind?}
\author{Yan Huang\altaffilmark{1,2} and Zhuo Li\altaffilmark{1,2}}
\affil{$^1$Department of Astronomy, School of Physics, Peking University, Beijing 100871, China; hyan623@pku.edu.cn\\
$^2$Kavli Institute for Astronomy and Astrophysics, Peking University, Beijing 100871, China \\
}
\begin{abstract}
The persistent soft X-ray emission from the location of the so-far most luminous supernova (SN), ASASSN-15lh (or SN 2015L), with $L \sim 10^{42} \unit{erg~ s^{-1}}$, is puzzling. We show that it can be explained by radiation from the SN-shock accelerated electrons inverse-Compton scattering the intense UV photons. The non-detection in radio requires strong free-free absorption in the dense medium. In these interpretations, the circumstellar medium is derived to be a wind ($n\propto R^{-2}$) with mass-loss rate of $\dot M  \ga 3 \times10^{-3}M_\odot(v_{\rm w}/10^3 \rm km\,s^{-1}) yr^{-1}$, and the initial velocity of the bulk SN ejecta is $\la 0.02c$.  These constraints imply a massive ejecta mass of $\ga60(E_0/2\times10^{52}\unit{erg})M_\odot$ in \asn, and a strong wind ejected by the progenitor star within $\sim8 (v_{\rm w}/10^3 \rm km\,s^{-1})^{-1}$ yrs before explosion.
\end{abstract}

\keywords{stars: mass-loss - supernovae:general - SLSN: individual (ASASSN-15lh)}

\section{Introduction}
Super-luminous supernovae (SLSNe) are a type of stellar explosions with a luminosity 10 or more times higher than the standard supernovae \citep[SNe;][]{gal12}. SN 2005ap was the first discovered SLSN, with an absolute magnitude at peak around $-22.7$ mag \citep{qui07}. Over the past decade, due to the large field-of-view, rapid-cadence transient searches, more than a hundred SLSNe had been found \citep{qui07, smi07, gal12, nic14}. SLSNe are likely associated with the deaths of the most massive stars, but the progenitors and the physics of the explosion are still not understood. Several power-input mechanisms have been proposed for SLSNe, including gamma-ray heating by the radioactive decays of $^{56}$Ni and $^{56}$Co \citep{gal09}, magnetar spin-down \citep{kas10, woo10}, SN shock interaction with dense material in the environment \citep{smi07, che11}, and the pair-instability SNe \citep{woo07,woo17}. Unveiling the progenitors and explosion mechanisms of SLSNe are crucial for our understanding of massive star evolution.

ASASSN-15lh was discovered by the All-Sky Automated Survey for Supernovae (ASAS-SN) with absolute magnitude of $-23.5$ and a peak bolometric luminosity of $L_{\rm bol}=(2.2 \pm 0.2) \times 10^{45} \unit{erg~s^{-1}}$, more than twice as previously known SLSNe, making it the most luminous SNe thus far \citep{don16}. The temporary behavior showed a unique double-humped structure. It reached the primary peak at several tens days after explosion, and then decayed. However, a rebrightening began $\simeq90$ days after the primary peak and was followed by a long plateau \citep{god17}. A persistent soft X-ray radiation with luminosity $L \sim \rm 10^{41} -10^{42} \unit{erg~ s^{-1}}$ at the location of ASASSN-15lh are observed by Chandra/CXO during the follow-up campaign \citep{mar17}. Radio follow-up of ASASSN-15lh had been carried out by ATCA, 197 days after first detection of optical observation, but no radio emission was detected \citep{koo15}.

The major features characteristics of ASASSN-15lh made it classified as a hydrogen-poor (type I) SLSN \citep{don16}\footnote{See, however, \cite{lel16,kruhler17}, who suggested that ASASSN-15lh was a tidal disruption event from a supermassive Kerr black hole.}.  There have been many discussions on the power input. \citet{don16} suggested that the large radiation energy does not favor the radioactive decay and magnetar as main energy sources, and the lack of spectral feature also disfavours the model of interaction with dense medium, however, \cite{met15, dai16} suggested that updated magnetar models may still work. The later observed rebrightening put new challenges to all the models. \citet{cha16} explained the double-humped structure as a signature of the interaction of massive SN ejecta of $\simeq 40 M_{\odot}$ with an H-poor circumstellar shell of $ \simeq 20 M_{\odot}$.

Here we focus on the origin of the X-ray emission, but not the UV emission, which is assumed to be produced from a region inside of the SN shock.
If the X-ray emission is really produced by \asn, one may expect it varying with time, other than keeping a constant luminosity, so the interpretation of X-ray emission from the host galaxy is favored \citep{mar17}. Here we show that the persistent behavior can be explained by radiation from the SN shock, i.e., the shock-accelerated electrons up-scattering the inner coming UV photons from the SN photosphere. In this way, the ejecta mass and the medium density can be derived, and gives hint of the progenitor of \asn. In \S 2, we give the main observational results of ASASSN-15lh. Our model is provided in \S 3, and \S 4 is results of modelling, followed by discussion and conclusion (\S 5). Notice that we use the $q_{x}=q/10^x$ convention and the cgs units in the following unless stated otherwise.

\section{Observations}

ASASSN-15lh was discovered on 2015 June 14 (UT) by the ASAS-SN survey \citep{don16}. The reshift is $z= 0.2326$, corresponding to a luminosity distance of $d_{\rm L}\simeq 1171 \unit{Mpc}$. At the primary peak of the light curve the absolute magnitude is $M_{\rm V, AB}= -23.5 \pm 0.1$, and the bolometric luminosity is $L_{\rm bol}= \rm (2.2\pm 0.2) \times 10^{45} erg~s^{-1}$. Because of similarities in temperature, luminosity and radius evolution between ASASSN-15lh and the other SLSNe-I, ASASSN-15lh was characterized as hydrogen-poor SLSNe-I \citep{don16}. The host galaxy of ASASSN-15lh is a luminous galaxy ($M_{\rm K} \simeq -25.5$) with little star formation.

The primary peak time is JD2457179 (June 05, 2015). \cite{god17} reported a UV rebrightening starting 90 days (observer frame) after the primary peak, was followed by a $\simeq 120$ days long plateau in the bolometric light curve, and faded again since $\simeq 210$ days after explosion. Over $\sim 550$ days since detection, ASASSN-15lh has radiated $ E_{\rm rad}\sim \rm 1.7-1.9 \times 10^{52} erg$.

\citet{mar17} presented the detection of a persistent soft X-ray emission, with a luminosity $L\sim \rm 10^{41} -10^{42} erg~s^{-1}$, at the location of ASASSN-15lh by Chandra. They obtained 4 epochs of deep X-ray observations with the Chandra/X-ray Observatory(CXO) on November 12, 2015, December 13, 2015, February 20, 2016, and August 19, 2016, respectively, corresponding to 129.4 days, 154.6 days, 201.5 days and 357.8 days (rest frame), respectively, since optical maximum light on June 5, 2015. Table \ref{Tab:Xray} shows the X-ray flux observed by CXO.

\begin{deluxetable}{ccccc}
\tabletypesize{\scriptsize}
\tablecolumns{8}
\tablewidth{0pc}
\tablecaption{Observed X-ray (0.3-10 keV) flux from \asn, assuming a power-law spectrum with photon index $\Gamma=3$ \citep{mar17}.}
\tablehead{ \colhead{Date (MJD)} &
\colhead{Exposure (ks)} & \colhead{Unabsorbed Flux ($\rm{erg~s^{-1}~cm^{-2}}$)}}
\startdata
57338 &  10  & $<2.0\times 10^{-15}$  \\
57369 & 10  & $\simeq4.4\times 10^{-15}$ \\
57438 &  40  & $\simeq3.6\times 10^{-15}$  \\
57619 &  30  & $\simeq4.9\times 10^{-15}$
\enddata
\label{Tab:Xray}
\end{deluxetable}

\citet{koo15} used ATCA to observe the radio emission, on November 21.1 UT, 2015, i.e., 197 days after the first detection of optical observation (MJD 57150.5). No radio emission was detected at the SN location. A 3-$\sigma$ upper limit of $\rm 23 \mu Jy$ at 5.5 GHz and $\rm 21 \mu Jy$ at 9 GHz were given.

\section{Model}
\subsection{Hydrodynamic evolution}
Consider that the SN ejecta of \asn~ drive a shock propagating into the circumstellar medium (CSM). The hydrodynamic evolution of the shock depends on the density structure of the freely expanding SN ejecta and the structure of the CSM. Consider the CSM to be a steady stellar wind released from the progenitor star of \asn. For a free wind with constant mass loss rate $\dot{M}$ and wind speed $v_{\rm w}$, one has the wind density as function of the radius $R$
\begin{align}
n=\frac{\dot{M}}{4 \pi R^2 m_{\rm p} v_{\rm w}}\equiv A R^{-2}.
\label{n}
\end{align}

So $A$ is a parameter representing the density of the wind-like CSM. For a high wind mass-loss rate of $ \dot{M}=\rm 10^{-2} M_{\odot} yr^{-1}$ and a wind speed of $ v_{\rm w}=10^8v_{\rm w,8}\rm cm\,s^{-1}$, $A$ should be normalized as $A=10^{38}A_{38}\rm cm^{-1}$.

Consider a spherical SN ejecta to be homogeneous with a constant velocity $\beta_0c$ and a bulk kinetic energy $E_0$. Initially the shock expands with the initial velocity $\beta=\beta_0$, transferring the ejecta energy into the swept-up medium. The shock energy when the shock expands to radius $R$ is
\begin{align}
E=(\beta c)^2\int_{0} ^R n m_{\rm p} 4 \pi r^2 dr
= 4 \pi A m_{\rm p} R (\beta c)^2.
\label{E}
\end{align}

Later on, when a half of the initial energy is transferred into the shocked medium, $E = \frac{1}{2} E_{0}$, the shock starts to decelerate significantly. This occurs at a radius defined as the deceleration radius,
\begin{align}
R_{\rm dec}=\frac{E_{\rm 0}}{8\pi m_{\rm p} A c^2 \beta_{\rm 0}^2} \approx 2.7 \times 10^{19} \unit{cm}~E_{\rm 0, 52}  A_{\rm 38}^{-1} \beta_{\rm 0,-2}^{-2}.
\label{Rdec}
\end{align}

For \asn, we normalize the initial energy to be $E_0=10^{52}E_{0,52}$, since the radiated energy is order of $10^{52}$erg. The corresponding deceleration time since the SN explosion is
\begin{align}
t_{\rm dec}=\frac{R_{\rm dec}}{c \beta_{\rm 0}} \approx 2.8\times10^3 \unit{yr}~E_{\rm 0, 52} A_{\rm 38}^{-1} \beta_{\rm 0,-2}^{-3}.
\label{tdec}
\end{align}

Thus, at time $t \leq t_{\rm dec}$, the shock propagates in a constant velocity, $\beta=\beta_{\rm 0}$, and the shock radius is
$R=c\beta_{\rm 0}t$. At $t>t_{\rm dec}$, the shock dynamics transits into the self-similar Sedov-Taylor solution, then we have the shock velocity $\beta=\beta_{\rm 0}^{2/3}(ct/R_{\rm dec})^{-1/3}$. Note, the sensitive dependence of $t_{\rm dec}$ on $\beta_0$ implies that in our case of non-relativistic shock in \asn, $\beta_0\ll1$, $t_{\rm dec}$ is much larger than the relevant observation time. So we only need to consider the free expanding phase of $t<t_{\rm dec}$ where $\beta=\beta_0$.

It should be noted that in a more realistic model one may consider a uniformly expanding ejecta with a steep power law density profile on the outside, for which the numerical simulations show that the outflow energy is as a function of velocity, $E_{\rm ej}(>\beta)\propto \beta^{-k}$. Since the shock energy is provided by the kinetic energy of the ejecta that catch up with the shock, the dynamical evolution is determined by $E_{\rm ej}(>\beta)=E(\beta)$, which gives $\beta\propto t^{1/(k+3)}$. Numerical simulations show that the velocity profile of the SN ejecta is $\beta \propto M_{\rm ej}(>\beta)^{-\beta_1 n/(n+1)}$, with $\beta_1\sim1/5$, and $n=3(3/2)$ for radiative (convective) envelopes of the progenitor stars \citep{mat99}, thus $k\sim14/3(19/3)$ correspondingly. The large $k$ values make $\beta=$constant a good assumption.

\subsection{Shock radiation}
Given the hydrodynamic evolution of the SN shock, we next discuss the radiation from the shock. The swept-up CSM electrons will be accelerated by the shock, via, e.g., diffusive shock acceleration processes, and the post-shock magnetic field is also amplified, hence the accelerated electrons will give rise to synchrotron and inverse-Compton (IC) radiation in the downstream region \citep[e.g.,][]{che06}. We discuss the IC and synchrotron components separately below, focussing on their contribution on X-ray and radio emission, respectively.

\subsubsection{IC radiation}
We first show that for \asn, IC is the dominant cooling process other than synchrotron radiation for the accelerated electrons. The synchrotron cooling time of electrons with Lorentz factor (LF) $\gamma_{\rm e}$ is $t_{\rm syn}=3m_{\rm e} c/4 \sigma_{\rm T} U_B \gamma_{\rm e}$, depending on the energy density of the post-shock magnetic field, $U_B=4\epsilon_B n m_{\rm p} (\beta c)^2$, where $\epsilon_B$ is the equipartition parameter for magnetic field. On the other hand, the electrons will also lose energy by up-scattering the ambient photons. Given the intense UV photon emission from the inner photosphere of the newly exploded SN, a dominant contribution of the seed photons for IC scattering is the UV photons \citep{bjo04}. For a UV luminosity of $L_{\rm UV}$, the photon energy density at the shock region is $U_{\rm ph}=L_{\rm UV}/4 \pi R^2 c$, then the electron cooling time due to IC scattering UV photons can be derived as $t_{\rm IC}=3m_{\rm e} c/4 \sigma_{\rm T} U_{\rm ph} \gamma_{\rm e}$. So the ratio of synchrotron to IC cooling time is
\begin{align}
\frac{t_{\rm syn}}{t_{\rm IC}}=\frac{U_{\rm ph}}{U_B} \approx 4.4 \times 10^2 L_{\rm UV, 45} \epsilon_{B,-1}^{-1} A_{\rm 38}^{-1} \beta_{0,-2}^{-2}.
\label{ratio}
\end{align}

For \asn, given the large UV luminosity \citep{god17} we normalize the UV luminosity as $L_{\rm UV}=10^{45}L_{\rm UV, 45}$erg. For $\beta\ll1$ and a wide range of $A$, one has $t_{\rm syn}\gg t_{\rm IC}$, hence we assume IC cooling dominates synchrotron cooling.

By diffusive shock acceleration theory, the CSM electrons swept-up by the SN shock are accelerated to follow a power law in momentum $dN_{\rm e}/dp_{\rm e}\propto p_{\rm e}^{-p}$ with $p_{\rm e}\geq p_{\min}$ and $p$ the power law index. For relativistic electrons we have $p_{\rm e}\propto\gamma_{\rm e}$, thus the electron distribution at $\gamma_{\rm e}\ga2$ can also be approximated as a power law in electron's LF with the same index, $dN_{\rm e}/d\gamma_{\rm e}\propto \gamma_{\rm e}^{-p}$. Radio observations of Type Ib/c SNe indicate $p \approx 3$ \citep{che06}, thus we take $p=3$ here. This is also consistent with the poorly constrained X-ray spectrum of \asn~ \citep{mar17}.

Define that the accelerated electrons carry a fraction $\epsilon_{\rm e}$ of the post-shock internal energy $U=n_{\rm e}m_{\rm p}\beta^2c^2$, with $n_{\rm e}$ the postshock proton number density. We will take the typical value $\epsilon_{\rm e}=0.1$. If the bulk electrons are relativistic, then using the approximation of power-law in LF, the minimum LF can be derived to be $\gamma_{\rm m}=[(p-2)/(p-1)](m_{\rm p}/m_{\rm e}) \epsilon_{\rm e} \beta^2$ \citep[e.g.,][]{pir13}. The minimum LF is a constant initially when the shock does not decelerate, $\beta=\beta_0$. However, if the initial shock velocity $\beta_0$ is low enough, the bulk electrons may be non-relativistic, $\gamma_{\min}-1\la1$. Since we are only interested in the electrons that emit synchrotron and IC radiation, the relevant electrons should be relativistic, $\gamma_{\rm e}\ga2$. Thus, for the characteristic frequencies in the synchrotron and IC spectra, we should take
\begin{equation}
  \gamma_{\rm m}=\max\left(\frac{p-2}{p-1}\frac{m_{\rm p}}{m_{\rm e}} \epsilon_{\rm e} \beta^2,2\right).
\end{equation}

If the bulk accelerated electrons are non-relativistic, the electron energy is, for $p\leq3$, dominated by electrons with $\gamma_{\rm e}\ga2$ \citep{sironi13}, i.e., relativistic electrons.
Thus, we can approximate $\epsilon_{\rm e} U\approx\int_2\gamma_{\rm e}m_{\rm e}c^2(dn_{\rm e}/d\gamma_{\rm e})d\gamma_{\rm e}$. Moreover, the postshock electron number density for $\gamma_e\geq2$ is $n_{\rm rel}\approx\int_2(dn_{\rm e}/d\gamma_{\rm e})d\gamma_{\rm e}$. Combining these two equations gives the fraction of relativistic, $\gamma_{\rm e}\geq2$, electrons in the total electrons,
\begin{equation}
  f_{\rm rel}\equiv\frac{n_{\rm rel}}{n_{\rm e}}=\min\left(1,\frac{p-2}{p-1}\frac{m_{\rm p}}{m_{\rm e}}\frac{\epsilon_e\beta^2}2\right).
\end{equation}

So given the total number of the shock swept-up electrons $N_{\rm e}=4 \pi A R$, the relativistic electron number that give rise to synchrotron and IC radiation is only $f_{\rm rel}N_{\rm e}$.

The radiative cooling changes the electron distribution. Let the electron cooling time, dominated by IC cooling, being equal to the age of the SN shock, $t_{\rm IC}=t$, we obtain the cooling LF $\gamma_{\rm c}=3m_{\rm e} c/4 \sigma_{\rm T} U_{\rm ph} t$. For electrons with $\gamma_{\rm e}>\gamma_{\rm c}$ the electrons cool significantly and the distribution deviates from the injection power law, with the index changed to $p+1$. Given the bright UV emission of \asn, we find that at the beginning $\gamma_{\rm c}<\gamma_{\rm m}$, all electrons are in fast cooling regime. Later $\gamma_{\rm c}>\gamma_{\rm m}$ may happen, thus we should consider both fast cooling and slow cooling regime in deriving the electron distribution and radiation spectrum.

Next we discuss the radiation spectrum from IC scattering UV photons, considering both fast and slow cooling cases like \citet{sar98}. On average, the IC radiation power of a single electron with $\gamma_{\rm e}$ is $P_{\rm IC}=(4/3)\sigma_{\rm T} c \gamma_{\rm e}^2 U_{\rm ph}$. For simplicity, we assume the seed photons are isotropic, neglecting the order of unity correction of anisotropic incoming photons. The UV photons are in a black body like spectrum, thus the energy distribution is narrow, and we can approximate them as monochromatic, with a frequency of $\nu_{\rm 0}=3kT_{\rm BB}/h$. Observations show thate the rest-frame temperature of the UV photons is $T_{\rm BB}\approx\rm 2.0 \times 10^4 K$ \citep{don16}.

Typically, the UV photons will be scattered by electrons with $\gamma_{\rm e}$ up to a frequency of $\nu_{\rm s}=(4/3) \gamma_{\rm e}^2 \nu_{\rm 0}$. The specific power at $\nu_{\rm s}$ is $P_{\rm m,IC} \approx P_{\rm IC}/\nu_{\rm s}=\sigma_{\rm T} c U_{\rm ph}/\nu_{\rm 0}$, independent of $\gamma_{\rm e}$. The relativistic electron number is $f_{\rm rel}N_{\rm e}$, and the observed IC flux at spectral peak is $F_{\rm m,IC}=f_{\rm rel}N_{\rm e} P_{\rm m,IC}/4 \pi d_{\rm L}^2 $, i.e.,
\begin{eqnarray}
F_{\rm m,IC} =163 \unit{\mu Jy} A_{\rm 38} L_{\rm UV,45}  t_{\rm d}^{-1} d_{\rm L, 28}^{-2} \beta_{\rm 0,-2}^{-1}f_{\rm rel},
\end{eqnarray}
where $t_{\rm d}=t/ (\rm 1~day)$. We take broken power law approximation for the radiation spectrum. The emergent IC spectrum \citep{ghi13} is, for fast cooling case ($\gamma_{\rm c}<\gamma_{\rm m}$)
\begin{equation}
F_{\rm \nu, IC}=
\begin{dcases}
F_{\rm m,IC} \left(\frac{\nu}{\nu_{\rm s,c}}\right),& \nu<\nu_{\rm s,c} \\
F_{\rm m,IC} \left(\frac{\nu}{\nu_{\rm s,c}}\right)^{-1/2},& \nu_{\rm s,c}  \leqslant \nu \leqslant \nu_{\rm s,m} \\
F_{\rm m,IC} \left(\frac{\nu_{\rm s,m}}{\nu_{\rm s,c}}\right)^{-1/2} \left(\frac{\nu}{\nu_{\rm s,m}}\right)^{-p/2}, & \nu>\nu_{\rm s,m}
\end{dcases}
\label{Ffast}
\end{equation}
whereas for slow cooling case ($\gamma_{\rm m}<\gamma_{\rm c}$),
\begin{equation}
F_{\rm \nu, IC}=
\begin{dcases}
F_{\rm m,IC} \left(\frac{\nu}{\nu_{\rm s,m}}\right),&\nu<\nu_{\rm s,m} \\
F_{\rm m,IC} \left(\frac{\nu}{\nu_{\rm s,m}}\right)^{-\frac{p-1}{2}},&\nu_{\rm s,m} \leqslant \nu \leqslant \nu_{\rm s,c} \\
F_{\rm m,IC} \left(\frac{\nu_{\rm s,c}}{\nu_{\rm s,m}}\right)^{-\frac{p-1}{2}} \left(\frac{\nu}{\nu_{\rm s,c}}\right)^{-p/2},&\nu>\nu_{\rm s,c}
\end{dcases}
\label{Fslow}
\end{equation}

The break frequencies in the spectrum are relevant to the characteristic electron LFs as
$\nu_{\rm s,c}=(4/3) \gamma_{\rm c}^2 \nu_{\rm 0}=1.4 \times 10^{7} \unit{Hz}~L_{\rm UV,45}^{-2}  t_{\rm d}^2 \beta_{\rm 0,-2}^4$,
and
$\nu_{\rm s,m}=(4/3) \gamma_{\rm m}^2 \nu_{\rm 0}=1.7 \times 10^{15} \unit{Hz}~\gamma_{\rm m}^2$.

\subsubsection{Synchrotron radiation}

Consider the synchrotron radiation by the shock-accelerated electrons, although it is not the dominant process of electron energy loss. In particular, we are interested in the synchrotron radiation contribution in the radio emission from SNe Ib/Ic \citep{che98, bjo04}. Define $\nu_{\rm a}$ as the frequency that synchrotron absorption optical depth is unity, and the electron LF that emitting photons at $\nu_{\rm a}$ as $\gamma_{\rm a}$. It should be noted, and confirmed later, that we are facing the problem with the minimum injection LF $\gamma_{\rm m}$ and the cooling LF $\gamma_{\rm c}$ being close to unity, and far smaller than $\gamma_{\rm a}$, i.e., $\gamma_{\rm a}\gg1$ and $\gamma_{\rm m}\sim\gamma_{\rm c}\sim1$. We still take broken power law approximation for the synchrotron spectrum.

The interested frequency range for GHz-radio emission would be the spectral segments around $\nu_{\rm a}$, which, as derived in \appendixname, is
\begin{align}
\nu_{\rm a} \approx 343 \unit{GHz}~ t_{\rm d}^{-\frac{p+3}{p+5}} L_{\rm UV,45}^{-\frac{2}{p+5}} A_{\rm 38}^{\frac{p+7}{2(p+5)}}  \epsilon_{B,-1}^{\frac{p+3}{2(p+5)}} \beta_{0,-2}^{\frac{2}{p+5}}f_{\rm rel}^{\frac{2}{p+5}}.
\label{nua}
\end{align}

No matter fast cooling $\gamma_{\rm c}<\gamma_{\rm m}$ or slow cooling regime $\gamma_{\rm m}<\gamma_{\rm c}$, the synchrotron flux at $\nu>\max(\nu_{\rm m},\nu_{\rm c})$ can be given by
\begin{equation}
F_{\nu, \rm syn}=
\begin{dcases}
F_{\rm m} \nu_{\rm m}^{(p-1)/2}\nu_{\rm c}^{1/2}\nu_{\rm a}^{-(p+5)/2}\nu^{5/2},&\nu<\nu_{\rm a}\\
F_{\rm m}  \nu_{\rm m}^{(p-1)/2}\nu_{\rm c}^{1/2}\nu^{- p/2},& \nu \geq \nu_{\rm a}
\end{dcases}
\label{syn}
\end{equation}
where $\nu_{\rm m}$ and $\nu_{\rm c}$ are the characteristic frequencies emitted by electrons with LFs of $\gamma_{\rm m}$ and $\gamma_{\rm c}$, respectively, and $F_{\rm m}=f_{\rm rel}N_{\rm e} P_{\rm m} /4 \pi d_{\rm L}^2$, with $P_{\rm m}=\sqrt{3} e^3 B/m_{\rm e} c^2$ being the synchrotron specific power of a single electron at its characteristic frequency $\nu=3 \gamma_{\rm e}^2 e B/ 4 \pi m_{\rm e} c$. Thus we can derive
\begin{equation}
  F_{\rm m}  = 2.87 \times 10^5  \unit{\mu Jy} A_{\rm 38}^{3/2} \epsilon^{1/2}_{B,-1}  d^{-2}_{\rm L,28} \beta_{0,-2}f_{\rm rel}.
\end{equation}

If the CSM is dense and ionized or partially ionized, free-free absorption could be important for radio emission. The optical depth of the wind from radius $R$ toward observer, due to free-free absorption, is given by
\begin{align}
\tau _{\rm ff} \approx 1.6 \times 10^8 T_{\rm e,4}^{-1.35} A_{\rm 38}^2 \beta_{0,-2}^{-3}\nu_{\rm 10}^{-2.1} t_{\rm d}^{-3},
\label{tff}
\end{align}
where $T_{\rm e}$ is the temperature of the electrons in the CSM  \citep{lan99}. We take $T_{\rm e}=2.0 \times 10^{4} \rm K$ below. The observed synchrotron flux after correction for free-free absorption should be $F_{\rm \nu}=F_{\nu, \rm syn} \exp[-\tau_{\rm ff}(\nu)]$.

\section{Parameter constraints}
\begin{figure}
\vskip -0.0 true cm
\centering
\includegraphics[scale=0.6]{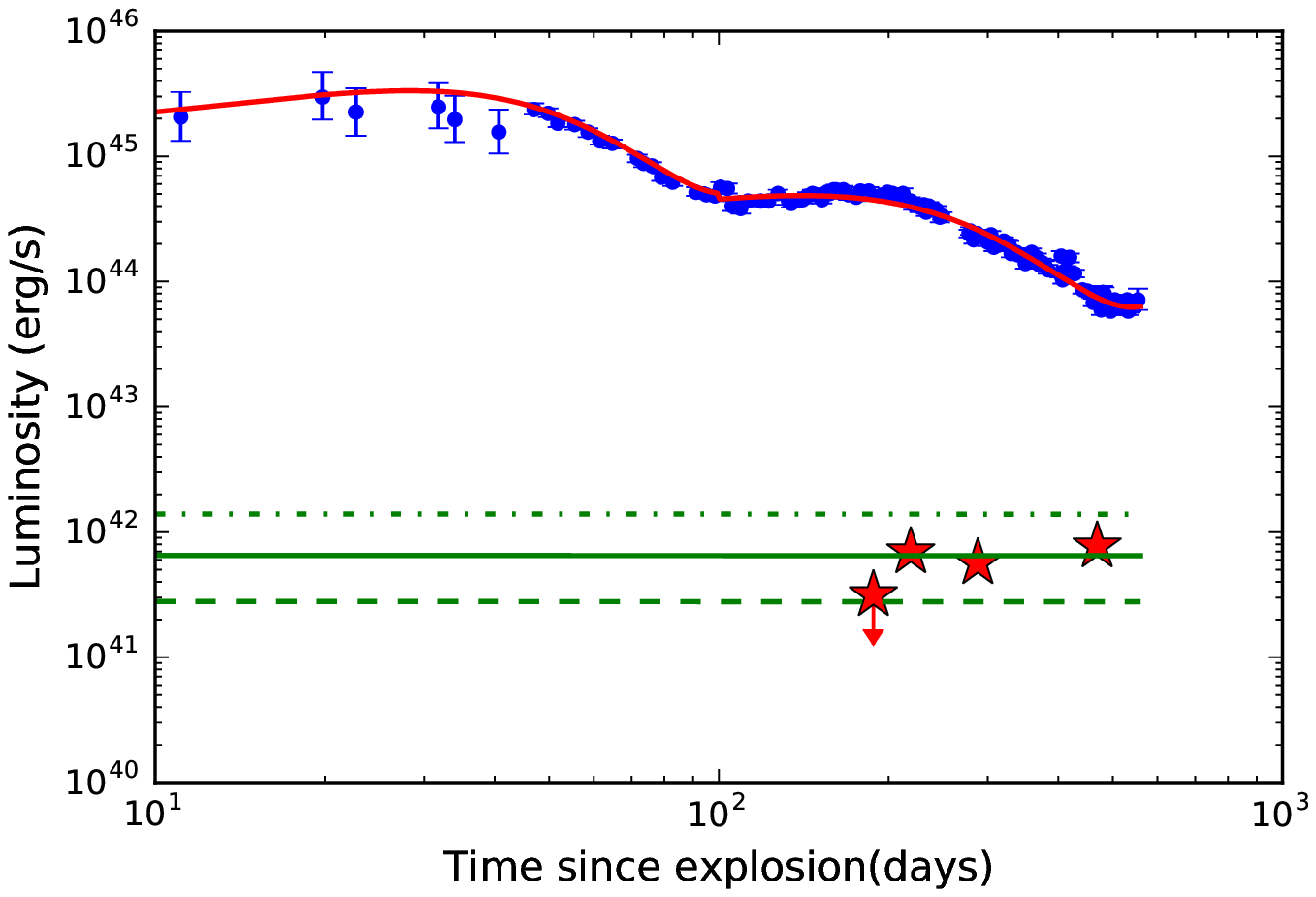}
\includegraphics[scale=0.6]{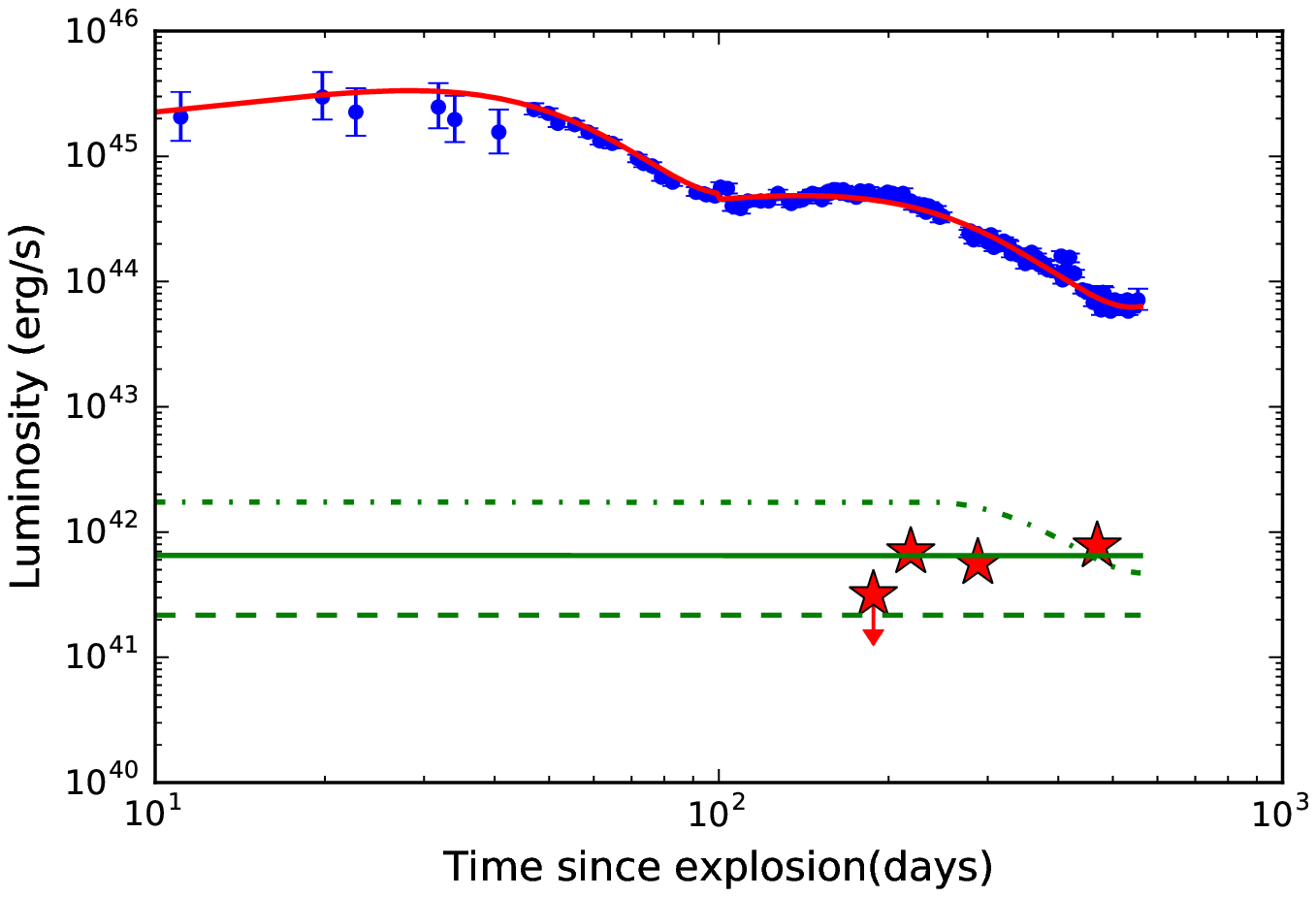}
\caption{IC radiation light curve from upscatterings of UV photons in comparison with observations. The upper panel is for fixing $\beta_{0,-2}=3$ but changing $A$: $A_{38}=$0.3, 0.7, and 1.5 for dashed, solid and dashed-dotted lines, respectively. The lower panel is for fixing $A_{38}=0.7$ but changing $\beta_0$: $\beta_{0,-2}=$1, 3, and 8 for the green dashed, solid and dashed-dotted lines, respectively. The red stars are the X-ray data. The blue dots show the bolometric luminosity evolution, and the red solid line shows the fitting function. The other parameters used are $\epsilon_{\rm e}=0.1$ and $p=3$.}
\label{Fig:LC}
\end{figure}
We use the model described above to fit the X-ray and radio data of \asn. The X-ray data can be interpreted by IC radiation due to upscattering of the intense UV radiation. In order to calculate the IC radiation we need the UV light curve as input, thus we first fit the UV light curve with two connecting third-order polynomials, $L_{\rm UV}=a_3 t^3 + a_2 t^2 + a_1 t + a_0$, with $(a_3,a_2,a_1,a_0)=(3.95 \times 10^{-6} , -7.77 \times 10^{-4} , 3.44 \times 10^{-2},11.5)$ for $t<100$ days (rest frame time since explosion), and $(2.77 \times 10^{-8},-2.83 \times 10^{-5} , 6.30 \times 10^{-3} ,10.7)$ for $t \geq 100$ days.
Here $L_{\rm UV}$ is in unit of $L_\odot$, and $t$ in day. This fitting function is shown in Fig. \ref{Fig:LC} as a red solid curve, in comparison with the UV data from \citet{god17}. With these seed photons, the calculated IC flux is also shown in Fig. \ref{Fig:LC}. We have integrated the IC flux over the energy range of $0.3-10$ keV to match the observed X-ray energy range. The IC luminosity is constant with time, well fitting the detected persistent X-ray flux (Table \ref{Tab:Xray}).

We explain why the IC flux is constant here. The X-ray emitting electrons are cooling fast due to upscattering the intense UV photons, so the X-ray range lies in a regime of $\nu\gg\max(\nu_{\rm s,m}, \nu_{\rm s,c})$. In this regime, no matter the bulk electrons are fast or slow cooling (i.e., $\gamma_{\rm c}<\gamma_{\rm m}$ or $\gamma_{\rm m}<\gamma_{\rm c}$), the IC flux is given by, see eqs. (\ref{Ffast}) and (\ref{Fslow}), $F_{\rm \nu, IC}=F_{\rm m,IC} \nu_{\rm s,m}^{(p-1)/2} \nu_{\rm s,c}^{1/2} \nu^{-p/2}$. During the evolution stage concerned, the SN-shock swept-up CSM material is not enough to decelerate the shock, and the shock keeps a constant velocity $\beta\simeq\beta_0$, hence the postshock electrons' characteristic LF is also a constant, since $\gamma_{\rm m}\propto\beta^2$ or $\gamma_{\rm m}=2$. Thus $ \nu_{\rm s,m}\propto\gamma_{\rm m}^2$ is a constant. Next, the total number of swept-up CSM electrons $N_{\rm e}\propto R\propto t$, within which the fraction of relativistic electrons is also constant, $f_{\rm rel}=1$ or $\propto\beta^2$, and the peak specific IC power is $P_{\rm m,IC}\propto U_{\rm ph}$, then the peak IC flux scales as $F_{\rm m,IC}\propto f_{\rm rel}N_{\rm e}P_{\rm m,IC}\propto U_{\rm ph}t$. At last, the electron cooling LF $\gamma_{\rm c}\propto U_{\rm ph}t$, thus $\nu_{\rm s,c}\propto\gamma_{\rm c}^2 \propto U_{\rm ph}^{-2} t^{-2}$. Putting all together we have $F_{\rm \nu, IC}\propto t^0$, being constant. In short, for a medium density with $n\propto R^{-2}$ and IC radiation in the fast cooling regime, the IC flux is a constant. Actually, for $\nu>\max(\nu_{\rm s,m},\nu_{\rm s,c})$ we can derive
\begin{equation}\label{eq:ICflux}
 F_{\nu, \rm IC}= 1.0 \times 10^{-6}  \unit{\mu Jy} A_{38} \beta_{\rm 0,-2} d_{\rm L, 28}^{-2} \gamma_{\rm m}^{p-1} \nu_{18}^{-p/2}f_{\rm rel},
\end{equation}
independent of time.

\begin{figure}
\vskip -0.0 true cm
\centering
\includegraphics[scale=0.6]{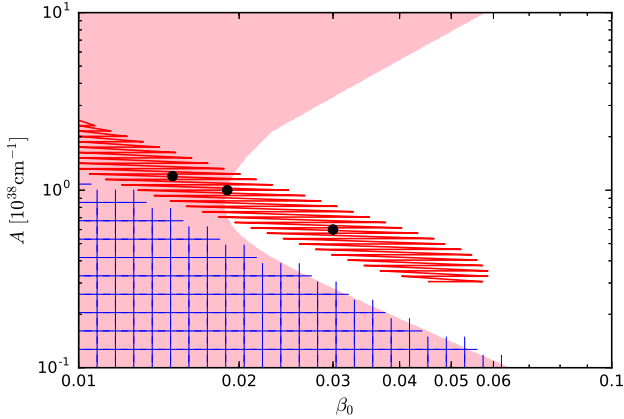}
\caption{The parameter constraints in the ($\beta_{0}, A$) 2D space with observations. The red oblique-line region shows the X-ray flux constraint. The blue grid (pink shaded) region shows the constraint from 9 GHz upper limit, without (with) free-free absorption taken into account. The parameters used are: $\epsilon_B=0.1$, $\epsilon_{\rm e}=0.1$, $p=3$ and $T_{\rm e}=\rm 2.0 \times 10^4K$. }
\label{Fig:parameter}
\end{figure}

The constant IC flux phase may end when the break $\nu_{\rm s,c}$ crosses the observation band. Letting $\nu_{\rm s,c}\simeq10^{18}$Hz we obtain that the crossing occurs at time
\begin{equation}
  t_{\rm cross}= 2.7\times10^5 \unit{days}~ L_{\rm UV, 45} \beta_{0,-2}^{-2}.
\end{equation}

After $t_{\rm cross}$ the X-ray band enters the regime of $\nu_{\rm s,c}>\nu>\nu_{\rm s,m}$, where the IC-produced flux at $\sim10^{18}$Hz is $F_{\rm \nu, IC} \propto F_{\rm m,IC} \propto U_{\rm ph} t\propto L_{\rm UV}t^{-1}$, decreasing with time.

In order to constrain the parameters, we apply the least square fitting method to fit the persistent X-ray emission. We define $ \chi^2=\sum_{i=1}^{N} (F_{\rm thy}[i]-F_{\rm obs}[i])^2$, where $F_{\rm thy}$ is the theoretical value calculated by the IC radiation in Eq. (\ref{Ffast}) and Eq. (\ref{Fslow}), $F_{\rm obs}$ is the observed X-ray flux as presented in Table \ref{Tab:Xray}, and $N$ is the total data number. We look for the minimum $\chi^2$ value, $\chi^2_{\rm min}$, in the ($A$,$\beta_{0}$) 2D space, then constrain the parameters $\beta_{0}$ and $A$ in the 2D space by requiring $\chi^2<2.0 \chi^2_{\rm min}$. The resulted parameter region is showed in red in Fig. \ref{Fig:parameter}. We see that there is a correlation between the constrained $A$ and $\beta_0$ values. Actually by equating the IC flux at $\nu>\max(\nu_{\rm s,m},\nu_{\rm s,c})$ (eq.\ref{eq:ICflux}) and the observed X-ray flux we can derive the $A-\beta_0$ relation, $A_{38}  \simeq 0.018 \beta_{\rm 0, -2}^{-1}f_{\rm rel}^{-1}$. Moreover, the upper limit on $\beta_0$ can be obtained by requiring $t_{\rm cross}$ being larger than observation time of the last X-ray data, $\beta_0\la0.06$.

\begin{figure}
\vskip -0.0 true cm
\centering
\includegraphics[scale=0.6]{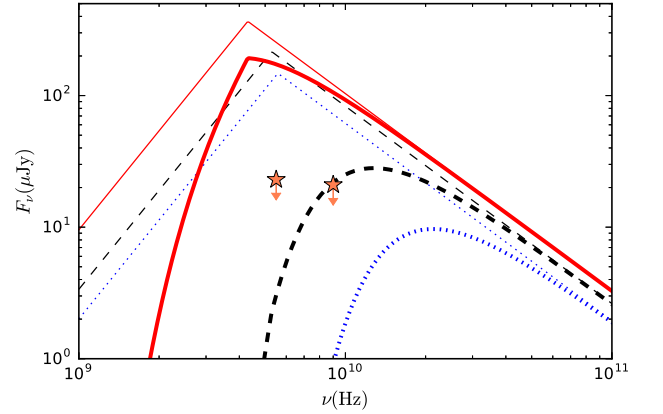}
\caption{The model synchrotron spectrum in comparison with the radio limits from observations. The thick (thin) lines represent synchrotron flux with (without) free-free absorption taken into account. Three cases satisfying X-ray constraints are, as marked as black solid points in Fig. \ref{Fig:parameter}, $(A_{38},\beta_{\rm 0,-2})=(0.6, 3)$ (red solid lines), $(1.0, 1.9)$ (black dashed lines), and $(1.2, 1.5)$ (blue dotted lines), respectively. The other parameters used are: $\epsilon_B=0.1$, $\epsilon_{\rm e}=0.1$, $p=3$ and $T_{\rm e}=\rm 2.0 \times 10^4K$.}
\label{Fig:Radio}
\end{figure}

The radio upper limits can further help to constrain parameters. Fig \ref{Fig:Radio} shows the synchrotron spectrum at 197 days in comparison with radio limits from observations. By requiring the synchrotron flux (eq.\ref{syn}) to satisfy the upper limit $F_{\nu,\rm syn}< 21 \mu$Jy at $\rm 9 GHz$ at 197 days, we constrain the allowed region in the $(A,\beta_0)$ space (Fig \ref{Fig:parameter}). Furthermore, by requiring the observed flux $F_{\rm \nu}=F_{\nu, \rm syn} \exp[-\tau_{\rm ff}(\nu)]$, with free-free absorption taken into account, to satisfy the observed limit, the allowed parameter region is larger, as shown in Fig \ref{Fig:parameter}. It is seen that there is not overlapping between the X-ray constrained region and the synchrotron self absorption only constrained region in Fig \ref{Fig:parameter}. So it is important to note that strong free-free absorption of the wind is required to account for the radio upper limit in \asn.

Combining the constraints by X-ray and radio observations, the allowed parameter ranges are $A\ga10^{38}\unit{cm^{-1}}$, and  $\beta_{0} \la 0.02$, as shown in Fig. \ref{Fig:parameter} the overlapping region between the X-ray constrained region (red-oblique-lined) and the radio constrained region (pink-shaded).

\section{Conclusion and discussion}

We have investigated the persistent X-ray emission from the location of \asn, and found that it can be produced by the SN shock propagating in a dense wind $(n\propto R^{-2})$, where the shock-accelerated electrons emit the X-rays by upscattering the inner-coming UV photons from the SN photosphere. We also found that the non-detection in radio requires that the wind is dense enough so that free-free absorption of the wind is important.  With observation data we can constrain that the wind density parameter is $A \ga 10^{38}\rm cm^{-1}$, and that the SN shock's initial velocity is $v_{\rm sh} < 0.02c$. This $A$ value corresponds to a stellar-wind's mass-loss rate of $\dot M  \geq  3 \times 10^{-3}M_\odot v_{\rm w,8} \rm yr^{-1}$, assuming a wind velocity of $v_{\rm w}=10^8\rm cm\,s^{-1}$.  The constrained SN shock velocity is somewhat lower compared to the average among the radio SNe, $v_{\rm sh}/c \approx 0.07$ \citep[e.g.,][and references there in]{kam16}.

The upper limit to the shock velocity $v_{\rm sh} \la 0.02c$ leads to a constraint on the ejecta mass of $M_{\rm ej} \sim 2E_0 /v_{\rm sh}^2 \ga56(E_0/2\times10^{52}\unit{erg})M_\odot$. This extreme large ejecta mass implies a massive progenitor star of \asn, consistent with a pair-instability SN \citep[e.g.,][]{woo07,woo17}.

With the constraints, $A_{38} \beta_{0,-2}\sim0.018$, and $\beta_0 \la 0.02$, we can calculate the shock energy (eq.\ref{E}) at $t\sim$500 days, $E\la 2 \times 10^{49}$ erg, much smaller than the total radiation energy of \asn, $\sim10^{52}$erg, and the typical kinetic energy of normal SNe, $\sim10^{51}$erg. The shock radius  at $t\sim$500 days is $R=\beta_0ct \la 8 \times 10^{-3} $pc, within which the CSM mass is about $M=4 \pi A m_{\rm p} \beta_0ct \simeq 0.027 M_{\odot}$. This mass should be ejected by the wind of \asn's progenitor within a time of $R/v_{\rm w}\la 8 v_{\rm w,8}^{-1}$ yrs before the SN explosion. Note this CSM mass is about 3 orders of magnitude lower than that is derived by using the interaction model to interpret the UV emission \citep{cha16}.

Recent-year observations of SN spectra within days of explosion have leaded to discovery of narrow emission lines in the early spectra of various kinds of SNe, indicating dense CSM immediately surrounding the progenitor stars. \cite{gal-yam14} first reported detection of strong emission lines in a SN IIb's early spectrum, indicating a strong Wolf-Rayet-like wind with $\dot{M} \sim 10^{-2} M_\odot\,\unit{yr^{-1}}(v_{\rm w}/500\rm km\,s^{-1})$. More recently, \cite{Yaron17} observed narrow emission lines from a regular type II SN within $10$hr after explosion, implying a dense wind of $\dot{M}\sim3 \times 10^{-3}M_\odot\,\unit{ yr^{-1}}(v_{\rm w}/100\rm km\,s^{-1})$ ejected yrs before explosion. The rapid spectra obtained within 5 days of SN II explosion have leaded to detection of narrow emission lines in a significant fraction, $18\%$, of early spectra of SNe II \citep{khazov16}. These observations imply that dense winds may be common in core-collapse SNe. Our interpretation of X-ray emission from \asn~ may indicate that type I SLSNe are also surrounded immediately by a dense wind, ejected $\sim10$yrs before the SLSN explosion.

\acknowledgements{We thank an anonymous referee for helpful suggestions. We also thank Yuanpei Yang, Tianqi Huang, Alexander Kann, Giorgos Leloudas, Wenbin Lu, and Yunwei Yu for helpful comments and discussions. This work is supported by the NSFC (No. 11773003) and the 973 Program of China (No. 2014CB845800).}

\appendix
\renewcommand{\appendixname}{Appendix~\Alph{section}}
\section{Synchrotron self-absorption frequency}\

The radio band that is interested here is in the frequency regime of $\nu \gg \max(\nu_{c}, \nu_{m})$. The electrons responsible to the radio emission are, due to fast cooling, distributed as $dn_{\rm e}/d\gamma_{\rm e}=C\gamma_{\rm e}^{-(p+1)}$ at $\gamma_{\rm e}(\nu) \gg\max(\gamma_{c}, \gamma_{m})$, where $p$ is the index of injected electrons. Using $\int_{\min(\gamma_{\rm c}, \gamma_{\rm m})}^{\infty} (dn_{\rm e}/d\gamma_{\rm e}) d\gamma_{\rm e}=4nf_{\rm rel}$, we derive $C \approx 4 f_{\rm rel}\gamma_{c} \gamma_{m}^{p-1}n$. The absorption coefficient at $\nu$ is given by \citep{ryb79}
\begin{align}
\alpha_{\rm \nu}=\frac{p+3}{8\pi m_{\rm e}\nu^2} \int_{\gamma_{\rm obs}} P(\tilde{\gamma}_{\rm e},\nu) C \tilde{\gamma}_{\rm e} ^{-p-2} d\tilde{\gamma}_{\rm e},
\end{align}
where $P(\gamma_{\rm e},\nu)=P_{\rm m}(\nu/\nu_{\rm syn}(\gamma_{\rm e}))^{1/3}$ is the specific synchrotron power by an electron with $\gamma_{\rm e}$, $\nu_{\rm syn}=3 \gamma_{\rm e}^2 e B/4 \pi m_{\rm e} c$, and $\gamma_{\rm obs}=(4 \pi m_{\rm e} c\nu/3 e B)^{1/2}$. Thus we further derive
\begin{eqnarray}
\alpha_{\rm \nu}&=&\frac{\sqrt{3} e^3}{8\pi m_{\rm e}^2 c^2} \left(\frac{3e}{4 \pi m_{e} c}\right)^{-1/3} \frac{p+3}{p+\frac{5}{3}} C B^{2/3}  \nu^{-5/3} \gamma_{\rm obs}^{-p-(5/3)}\\
&=& 5.32 \times 10^{33} \unit{cm^{-1}} f_{\rm rel}t_{\rm d}^{-\frac{p+5}{2}} L_{\rm UV,45}^{-1} A_{\rm 38}^{\frac{p+7}{4}}  \epsilon_{B,-1}^{\frac{p+3}{4}} \nu^{-\frac{p+5}{2}},
\end{eqnarray}
where in the second equation we have plugged in the expressions for $C$, $B$ and $\gamma_{\rm obs}$, and the coefficient is calculated for $p=3$.
By setting the optical depth $\tau\approx\alpha_{\rm \nu} R/10 = 1$, we obtain the absorption frequency
\begin{align}
\nu_{\rm a} \approx 343 \unit{GHz}~ t_{\rm d}^{-\frac{p+3}{p+5}} L_{\rm UV,45}^{-\frac{2}{p+5}} A_{\rm 38}^{\frac{p+7}{2(p+5)}}  \epsilon_{B,-1}^{\frac{p+3}{2(p+5)}} \beta_{0,-2}^{\frac{2}{p+5}}f_{\rm rel}^{\frac{2}{p+5}}.
\label{va}
\end{align}

\end{document}